\documentclass[twocolumn,prd,superscriptaddress,nofootinbib,notitlepage]{revtex4-1}
\pdfoutput=1
\usepackage[T1]{fontenc}
\usepackage{bm}
\usepackage{comment}
\usepackage{subfigure}
\usepackage{amssymb}
\usepackage{graphicx}
\usepackage{amsmath}
\usepackage{tensor}
\usepackage{epstopdf}
\usepackage{extarrows}
\usepackage{xcolor}
\usepackage[colorlinks=true,
              linkcolor=blue,
              urlcolor=blue,
              citecolor=blue
              ]{hyperref}
\usepackage{latexsym}
\usepackage{amsmath,amsfonts,graphicx,accents}
\usepackage{amsbsy}
\usepackage{hyperref}
\usepackage{mathrsfs}
\usepackage{color}
\usepackage{psfrag}
\usepackage{enumerate}
\usepackage{amsmath,amssymb,calc,amsfonts}
\usepackage{latexsym}
\usepackage[utf8]{inputenc}
\usepackage[percent]{overpic}
\usepackage[autostyle=false, style=english]{csquotes}
\MakeOuterQuote{"}
\DeclareFontFamily{U}{rsfs}{}         
\DeclareFontShape{U}{rsfs}{m}{n}{<5> rsfs5 <6><7> rsfs7          %
  <8><9><10><10.95><12><14.4><17.28><20.74><24.88> rsfs10}{}     %
\DeclareMathAlphabet{\mathfs}{U}{rsfs}{m}{n}                     %
\definecolor{indiagreen}{rgb}{0.07, 0.53, 0.03}
\def\beq{\begin{eqnarray}}
\def\eeq{\end{eqnarray}}

\def\={\stackrel{\Delta}{=}}

\newcommand{\ave}[1]{\left<{#1}\right>}

\begin{document}
\title{Thermodynamics of Einstein-Gauss-Bonnet Black Holes and Ensemble-averaged Theory}
\author{Md Sabir Ali}\email{alimd.sabir3@gmail.com}
\affiliation{Department of Physics, Mahishadal Raj College, West Bengal, 721628 India}
\author{C. Fairoos}\email{fairoos.phy@gmail.com}
\affiliation{Department of Physics, T. K. M. College of Arts and Science Kollam, Kerala, 691005 India}
 \author{C. L. Ahmed Rizwan}\email{ahmedrizwancl@gmail.com}
\affiliation{Department of Physics, Kannur University, Payyanur, Kerala, 6710327, Kerala, India}
\author{T. K. Safir}\email{stkphy@gmail.com}
\affiliation{Department of Physics, T. K. M. College of Arts and Science Kollam, Kerala, 691005 India}
\author{Peng Cheng}\email{p.cheng.nl@outlook.com}
\affiliation{Center for Joint Quantum Studies and Department of Physics, School of Science, Tianjin University, Tianjin 300350, China}
\begin{abstract}
In this paper, using the ensemble-averaged theory, we define the thermodynamic free energy of Einstein-Gauss-Bonnet (EGB) black holes in anti-de Sitter (AdS) spacetime. This approach derives the gravitational partition function by incorporating non-saddle geometries besides the classical solutions. Unlike the sharp transition points seen in free energy calculated via saddle-point approximation, the ensemble-averaged free energy plotted against temperature shows a smoother behavior, suggesting that black hole phase transitions may be viewed as a small-$G_N$ (Newton’s gravitational constant) limit of the ensemble theory. This is similar to the behavior of black hole solutions in Einstein’s gravity theory in AdS spacetime. We have obtained an expression for the quantum-corrected free energy for EGB-AdS black holes, and in the six-dimensional case, we observe a well-defined local minimum after the transition temperature which was absent in the earlier analysis of the classical free energy landscape. Furthermore, we expand the ensemble-averaged free energy in powers of $G_N$ to identify non-classical contributions. Our findings indicate that the similarities in the thermodynamic behavior between five-dimensional EGB-AdS and Reissner-Nordström-AdS (RN-AdS) black holes, as well as between six-dimensional EGB-AdS and Schwarzschild-AdS black holes, extend beyond the classical regime.

\end{abstract}
  
\maketitle
\section{Introduction}

The thermodynamics of black holes in asymptotically AdS spacetime holds a unique position within the study of quantum gravity, as it interlaces principles from gravitation, thermodynamics, and quantum mechanics.  A notable example is the AdS/CFT correspondence, a powerful framework that enables the study of gravitational systems via dual quantum field theories. Within this framework, the Hawking-Page phase transition—a first-order transition between radiation and black hole states in Schwarzschild-AdS black holes immersed in a thermal bath \cite{Hawking:sw}—finds a dual interpretation. This transition corresponds to the confinement/deconfinement phase transition in a quark-gluon plasma within the boundary quantum field theory \cite{Witten:1998zw}. Such insights from AdS/CFT have significantly expanded our understanding of quantum gravity and provide an avenue to address the black hole information paradox. Furthermore, the negative cosmological constant characterizing the AdS spacetime gives rise to complex phase structures, a field now known as black hole chemistry \cite{Kubiznak:2016qmn}. Through this framework, the holographic duality extends from black holes to conformal field theories (CFT) \cite{Maldacena:1997re, Witten:1998qj}, to phenomena in quantum chromodynamics \cite{Kovtun:2004de}, and even to condensed matter systems, particularly those with strong coupling \cite{Hartnoll:2008vx, Hartnoll:2007ih}.\\

When a black hole is treated as a thermal system, drawing analogies with conventional thermodynamics in any diffeomorphism-invariant gravity theory is straightforward. However, the microscopic description of the black hole event horizon, and the related phase transition behavior is formidable. Despite extensive progress in black hole thermodynamics, the statistical foundation of these phase transitions remains elusive. Although the thermal properties of black holes in classical geometry are well understood, developing a full statistical framework is ongoing. The Euclidean path integral method, with a partition function approach, presents a promising avenue \cite{Gibbons:1976ue, Gibbons:1978ac, Hawking:1978jz}. Yet, integrating quantum gravity effects within this framework remains a complex task. In the path integral formulation of gravity, the saddle point approximation identifies action extrema, corresponding to the global minimum of the on-shell Euclidean action. While on-shell geometries provide insights into classical black hole thermodynamics, off-shell geometries must be considered to fully capture the subtleties of black hole phase transitions. Recent work has explored the ensemble-averaged description of black hole thermodynamics by including non-classical contributions to the path integral \cite{Cheng:2024efw, Cheng:2024hxh}. Inclusion of non-saddle contributions to the partition function results modified expression for the free energy and one does not observe a sharp phase transition point. In this construction, the black hole phase transition is interpreted as the small $G_N$ limit of ensemble averaged theory. \\

In this paper, we investigate the thermodynamics of EGB-AdS black holes beyond the classical limit by applying the ensemble-averaged theory. EGB theory is the natural extension of general relativity with higher-curvature contributions \cite{Lovelock:1971yv}. Also, it presents a rich thermodynamic structure. Notably, the thermodynamic behavior of five-dimensional EGB black holes resembles that of Reissner-Nordström-AdS black holes, while six-dimensional EGB black holes share similarities with Schwarzschild-AdS black holes. However, these correspondences have been explored only within the classical limit. Here, we extend this analysis to examine whether these thermodynamic parallels persist beyond the classical framework.\\

The paper is organized as follows. In Sec.\ref{thermodynamics}, we examine the classical thermodynamics of black holes in a generic $D$-dimensional EGB-AdS spacetime. Sec.\ref{second} extends this analysis using the ensemble-averaged theory, evaluating the gravitational partition function by including non-saddle geometries, in contrast to the usual approach that relies on the saddle-point approximation. We present numerical results for the ensemble-averaged free energy of five- and six-dimensional EGB-AdS black holes for various values of $G_N$. In Sec.\ref{four}, we expand the free energy in powers of $G_N$ and identify the quantum corrections at subleading and sub-subleading orders. Finally, in Sec.\ref{dis}, we discuss our findings and conclude.

\section{Thermodynamics of EGB-AdS Black Holes in $D$ Dimensions}\label{thermodynamics}
We start with a brief overview of the charged AdS black holes in Gauss-Bonnet gravity theory. In this theory, we have the Einstein-Gauss-Bonnet action in the presence of a negative cosmological constant and Maxwell's electrodynamic field.
\begin{widetext}
\begin{eqnarray}
S=\frac{1}{16\pi G_N}\int d^D x\sqrt{-g}\left(R+\frac{(D-1)(D-2)}{L^2}+\alpha(R_{\mu\nu\alpha\beta}R^{\mu\nu\alpha\beta}-4R_{\mu\nu}R^{\mu\nu}+R^2)+\mathcal{L}(\mathcal{F})\right), 
\end{eqnarray}
\end{widetext}
where $L$ is the AdS radius, $\alpha$ ($\geq 0$) is the Gauss-Bonnet coupling constant, $\mathcal{L}(\mathcal{F})=-F^{\mu\nu}F_{\mu\nu}$ is the Maxwell Lagrangian, with $F_{\mu\nu}=2\nabla_{[\mu}A_{\nu]}$. The spacetime metric for $D$-dimensional, asymptotically AdS, EGB gravity in the presence of an electromagnetic gauge field is given by,
\beq
ds^2 = -f(r) dt^2+\frac{1}{f(r)} dr^2+r^2d\Omega_{D-2}^2,
\eeq
where $d\Omega_{D-2}$ is the metric of the $(D-2)$ dimensional sphere of unit radius.
 The metric function $f(r)$ for the black hole solution is given by\cite{Cai:2013qga},
 \small
\begin{eqnarray}
f(r) = 1+\frac{r^2}{2 \tilde{\alpha}}\Bigg[ 1-\sqrt{1+4 \tilde{\alpha} \left(\frac{m}{r^{D-1}}-\frac{q^2}{r^{2D-4}}-\frac{1}{L^2}\right)} \ \Bigg]
\label{metricfunction}
\end{eqnarray}
\\
where $\tilde{\alpha}=(D-3)(D-4)\alpha$. The parameters $m,q$ are related to the mass ($M$) and charge ($Q$) of the hole by the following relations:
 \small
\begin{eqnarray*}
M = \frac{(D-2) \Omega_{D-2}}{16 \pi G_N} \ m,\; Q = \sqrt{2(D-2)(D-3)}\;q,
\end{eqnarray*}


One can express the basic thermodynamic quantities associated with the black hole in terms of its event horizon ($r_h$) which is defined by $f (r_h)=0$. Accordingly, the mass, temperature, and the entropy of the black hole are given, respectively, as,
\begin{widetext}
\beq \label{Haw_T}
&&M=\frac{(D-2)\Omega_{D-2}}{16\pi G_N}\Bigg[\frac{r_h^{D-1}}{L^2}+r_h^{D-3}+\tilde{\alpha}r_h^{D-5}+\frac{q^2}{r_h^{D-3}}\Bigg],\\
&&T_H= \frac{1}{4\pi r_h (r_h^2+2 \tilde{\alpha})}\Bigg[(D-3) r_h^2 + \frac{D-1}{L^2} r_h^4+(D-5) \tilde{\alpha}-\frac{(D-3)}{r_h^{2(D-4)}} q^2\Bigg],\\  
&&S= \frac{1}{4G_N} \Omega_{D-2} r_h^{D-2} \left(1+\frac{2 \tilde{\alpha} (D-2)}{(D-4)} \frac{1}{r_h^2}\right).
\eeq 
\end{widetext}
In the canonical ensemble description of black hole thermodynamics, the black hole is considered in contact with a thermal bath at a fixed temperature $T$. To examine the phase-switching dynamics of the black hole due to thermal fluctuations, one can use the free energy landscape approach. In this framework, the state of the black hole is defined by taking the horizon radius as the system's order parameter. Each state is referred to as a fluctuating black hole, and phase transition occurs between two stable black hole states through a series of unstable fluctuating black hole states. The Hawking–Gibbons path integral method for quantum gravity offers an elegant way to derive the thermodynamic free energy of fluctuating black holes also called generalized free energy. In this approach, the partition function in the canonical ensemble theory of gravity is obtained by integrating the Euclidean action $(I_E[g])$ over all geometries, i.e.,
\beq \label{partition_full}
Z_{grav}(\beta) = \int D[g] \ e^{-I_E[g]},
\eeq
where $\beta=1/T$ is the period of Euclidean time. The Euclidean action for EGB gravity theory can be obtained as \cite{Li:2023men},
\begin{widetext}
\beq \label{IE_D}
I_E=\frac{(D-2)\Omega_{D-2} \beta}{16 \pi G_N}\Bigg[r_h^{D-3}+\frac{r_h^{D-1}}{L^2}+\tilde{\alpha} r_h^{D-5}+\frac{q^2}{r_h^{D-3}}\Bigg] - \frac{1}{4G_N}\Omega_{D-2}r_h^{D-2}\left(1+\frac{2 \tilde{\alpha} (D-2)}{(D-4)}\frac{1}{r_h^2}\right)
\eeq
\end{widetext}
In the standard free energy landscape description, one exploits the saddle-point approximation to obtain the gravitational partition function. This approximation is reasonable since the maximum contribution to the partition function comes from the classical geometry which describes the black hole. Once we have the partition function, it is straightforward to obtain the free energy from the relation $F = -T \ln Z_{grav}$. For the case of EGB theory, the free energy for the fluctuating black holes is given by, 
\begin{widetext}
\beq \label{free_D}
F=\frac{(D-2)\Omega_{D-2}}{16 \pi}\Bigg[r_h^{(D-3)}+\frac{r_h^{(D-1)}}{L^2}+\tilde{\alpha} r_h^{(D-5)}+\frac{q^2}{r_h^{(D-3)}} - \frac{4 \pi T}{(D-2)} r_h^{(D-2)}\left(1+\frac{2 \tilde{\alpha} (D-2)}{(D-4)}\frac{1}{r_h^2}\right)\Bigg]
\eeq   
\end{widetext}
The generalized free energy obtained represents a collection of black hole states, each associated with a different horizon temperature. Thermodynamically stable black hole states correspond to the local extrema of the generalized free energy function. In other words, when the ensemble temperature is equal to the Hawking temperature the black hole state is in equilibrium with the thermal bath. The generalized free energy is used to construct the free energy landscape and various thermodynamic properties of EGB black holes can be obtained\cite{Li:2023men} (and the references therein). It is important to note that all these black hole states are derived solely from the classical geometry. To understand the non-classical contributions to black hole thermodynamics we use the ensemble-averaged theory as proposed in \cite{Cheng:2024hxh}. In the following section, we discuss the ensemble-average theory for five and six-dimensional EGB-AdS gravity theory.

\section{Black Hole Thermodynamics and Ensemble Average Theory} \label{second}

As discussed in the previous section, the gravitational path integral formulation of black hole thermodynamics is, so far, limited to contributions from classical saddle points. In other words, the free energy expression was derived directly from the classical Euclidean action without performing the full integration. Ideally, the gravitational partition function should involve integrating over all possible Euclidean geometries, weighted by the Euclidean action $I_E[g]$, as outlined in Eq.~\eqref{partition_full}. Consequently, any physical quantity associated with the Lorentzian metric must be averaged across all ensembles. In this spirit, one can define the ensemble average of a physical quantity $A[g]$ by,

\beq
\ave{A} = \frac{\int A[g] \ e^{-I_E[g]} \ D[g]}{\int e^{-I_E[g]} \ D[g]}
\eeq

Here, the integration is over various geometries described by $D[g]$. In this context, $e^{-I_E[g]}$ can be interpreted as the probability of obtaining the value $A$ from the geometry $g_{\mu \nu}$, with the partition function in the denominator serving as a normalization factor. Since we are interested in examining the nature of phase transitions in the theory of gravity, we focus on the ensemble average of the free energy. The key question that arises is: what variables should be considered when performing the integration? A natural choice will be to describe the geometry in terms of the canonical coordinate $r_h$ and its conjugate momentum $\dot{r_h}$. This construction is the same as in the standard statistical mechanical description. However, in the case of gravity theory, integrating over $\dot{r_h}$ contributes a constant on both the numerator and the denominator, and therefore gets canceled. To this extent, the ensemble average of the free energy for the $D$- dimensional EGB-AdS gravity theory is given as,
\beq \label{ESBAVG5D}
\ave{F} = \frac{\int F(r_h) \ e^{-I_E(r_h)} \ dr_h}{\int e^{-I_E(r_h)} \ dr_h},
\eeq
where $I_E(r_h)$ and $F(r_h)$ are given in Eq. \eqref{IE_D} and Eq.  \eqref{free_D} respectively. As the thermodynamic behavior of a five-dimensional EGB black hole is different from a six-dimensional one, we examine the ensemble-averaged theory separately.

\subsection{ Case-I: Five dimensions }

\begin{figure*}[hbt!]
    \centering
\mbox{
    \subfigure[ Average free energy $\ave{F}$ for the five-dimensional neutral EGB black hole for  $G_N = 0.02$ and $1$. \label{fig_1_1}]{\includegraphics[width=0.4\textwidth]{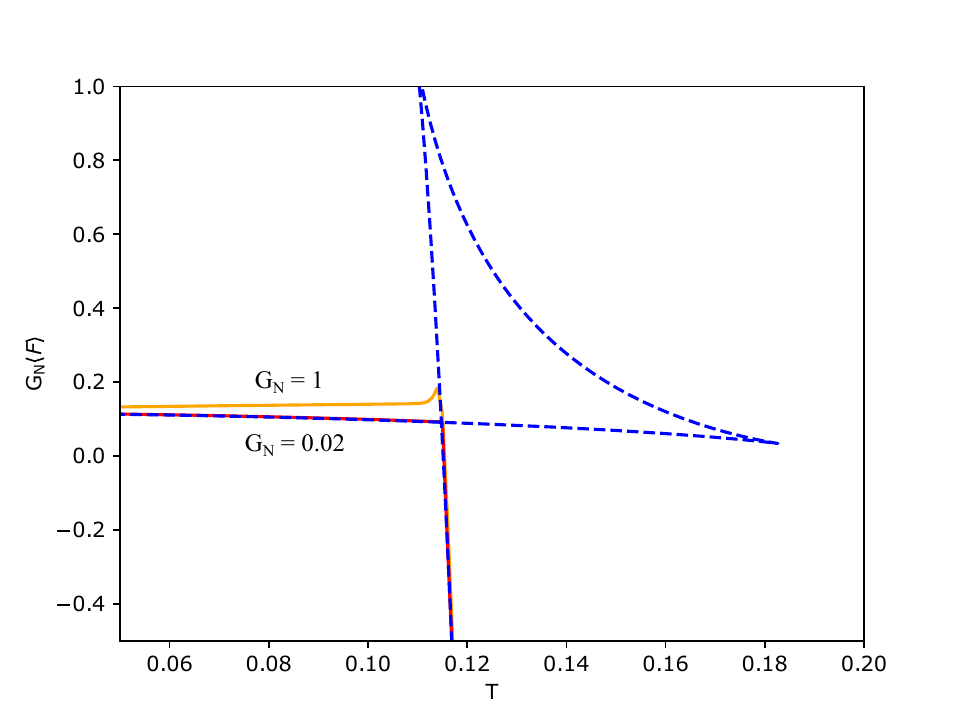}}\quad
    \subfigure[Average free energy $\ave{F}$ for the five-dimensional EGB spacetime in the presence of an electromagnetic gauge field for  $G_N = 0.02$ and $2$.\label{fig_1_2}]{\includegraphics[width=0.4\textwidth]{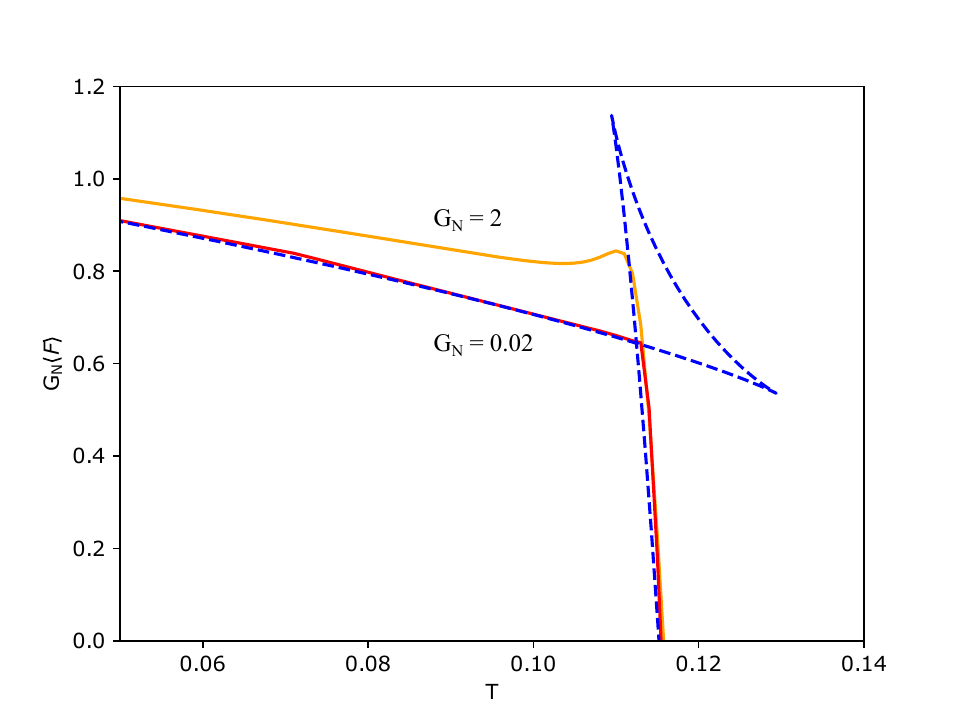}}\quad
  }
      \caption{As $G_N$ increases the sharp characteristics of phase transition are absent in both cases. The dashed blue curve represents the free energy obtained using the saddle point approximation.}
  \label{fig_1}
\end{figure*}

\begin{figure*}[hbt!]
    \centering
\mbox{
    \subfigure[ Average free energy $\ave{F}$ for the six-dimensional neutral EGB black hole for  $G_N = 0.01$ and $1$. \label{fig_2_1}]{\includegraphics[width=0.4\textwidth]{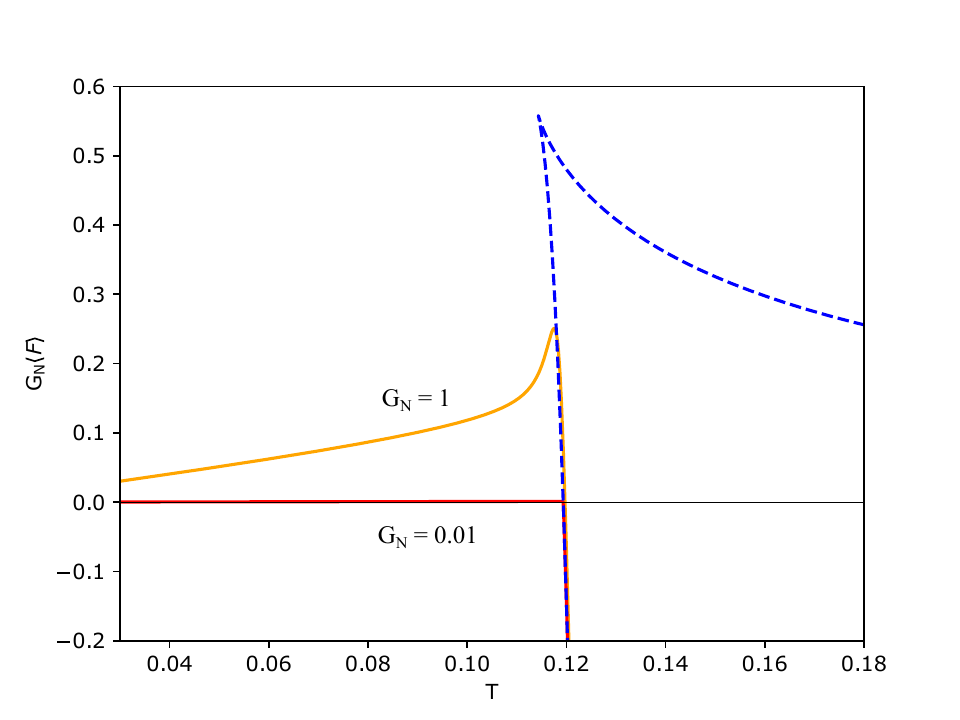}}\quad
    \subfigure[Average free energy $\ave{F}$ for the six-dimensional EGB spacetime in the presence of an electromagnetic gauge field for  $G_N = 0.01$ and $1$.\label{fig_2_2}]{\includegraphics[width=0.4\textwidth]{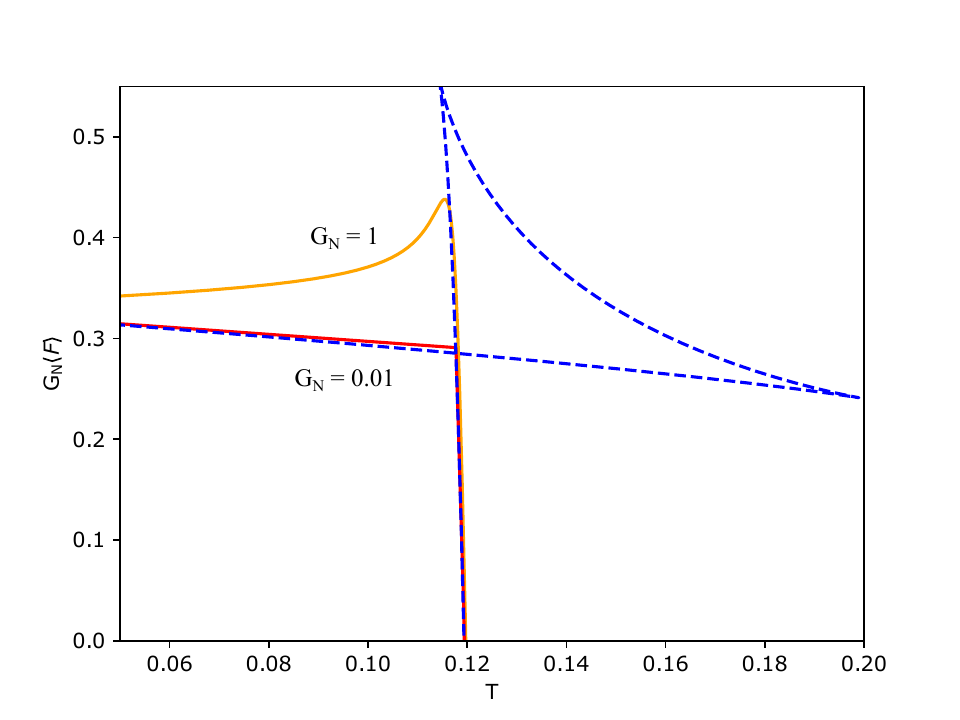}}\quad
  }
      \caption{As $G_N$ increases the sharp characteristics of phase transition is absent in both cases. The dashed blue curve represents the free energy obtained using the saddle point approximation.}
  \label{fig_2}
\end{figure*}
The Euclidean action for five-dimensional EGB gravity theory is given by \cite{Li:2023men},
\small
\beq
I_E = \frac{3 \pi \beta r_h^2}{8 G_N} \Big[ 1+\frac{r_h^2}{L^2}+\frac{\tilde{\alpha}}{r_h^2}+\frac{q^2}{r_h^4}-\frac{4 \pi}{3} T r_h\left(1+\frac{6 \tilde{\alpha}}{r_h^2}\right)\Big].
\eeq

The corresponding free energy is obtained by differentiating the Euclidean action with respect to $\beta$. Now, the ensemble average of free energy can be obtained numerically using Eq. \eqref{ESBAVG5D}. When $q=0$, the classical and ensemble-averaged free energies are plotted against the ensemble temperature in Fig. \ref{fig_1_1} for two distinct values of $G_N$. As observed in numerous studies, the free energy derived through the saddle-point approximation exhibits a distinct phase transition between small and large black holes, clearly indicated by the blue-dashed curve. This behavior resembles the RN-AdS spacetime and is supported by interpreting the Gauss-Bonnet coupling constant as a gauge field parameter. However, similar to the RN-AdS case, the ensemble average of free energy has no distinct turning point.
Furthermore, as $G_N$ approaches zero, the ensemble-averaged and the classical free energy curves merge. Thus we conclude that the black hole phase transition can be viewed as a small $G_N$ approximation of ensemble-averaged physics. In the presence of an electromagnetic gauge field, the deviation between the ensemble-averaged free energy and its saddle-point counterpart becomes more pronounced, as shown in Fig. \ref{fig_1_2}. These findings highlight the quantum mechanical nature of the ensemble system under study. For larger values of $G_N$, the likelihood of the system occupying states beyond the classical solution increases, a behavior marked by the absence of a sharp phase transition point between the small and large black hole phases.

\subsection{Case-II: Six dimensions }
In EGB gravity theory, the thermodynamic behavior of black holes in six dimensions differs from that in five dimensions. This distinction primarily arises from the nature of the Hawking temperature given in Eq. \eqref{Haw_T}. When $D=5$, one of the terms in Eq. \eqref{Haw_T} vanishes, impacting the free energy expression as seen in Eq. \eqref{free_D}. This specific term, however, contributes to the thermodynamic structure in all other dimensions. Consequently, for neutral black holes in six-dimensional EGB theory, a small-intermediate-large phase transition does not occur; rather, the phase transition resembles that of Schwarzschild-AdS spacetime, i.e., a transition between thermal AdS space and a black hole phase. In contrast, when an electric charge is present, the behavior of six-dimensional black holes closely aligns with that of five-dimensional ones.\\

The ensemble-averaged expression for the free energy includes contributions from non-classical geometries. In Fig. \ref{fig_2}, the behavior of free energy obtained from both the saddle-point approximation and the ensemble-averaged theory is plotted against the ensemble temperature for various values of $G_N$. The free energy from the saddle-point approximation shows a sharp phase transition, as indicated by the dashed blue lines. A key observation is the absence of a sharp turning point in the ensemble-averaged theory, similar to what is observed in the five-dimensional case. This characteristic is due to the non-classical contributions to the free energy, which become more pronounced as $G_N$ increases. For $q=0$, the ensemble-averaged free energy behaves similarly to the Schwarzschild-AdS black hole described in \cite{Cheng:2024hxh}. When an electromagnetic gauge field is present, the ensemble-averaged free energy qualitatively resembles the behavior seen in the five-dimensional EGB theory. The conclusion that the black hole phase transition can be understood as a small $G_N$ approximation of the ensemble-averaged theory applies to six-dimensional EGB gravity as well.

\section{Non-classical Corrections to the Black Hole Thermodynamics}
\label{four}
\begin{figure*}[hbt!]
    \centering
\mbox{
    \subfigure[The blue line represents the quantum corrected free energy of EGB black hole in 5D. \label{}]{\includegraphics[width=0.4\textwidth]{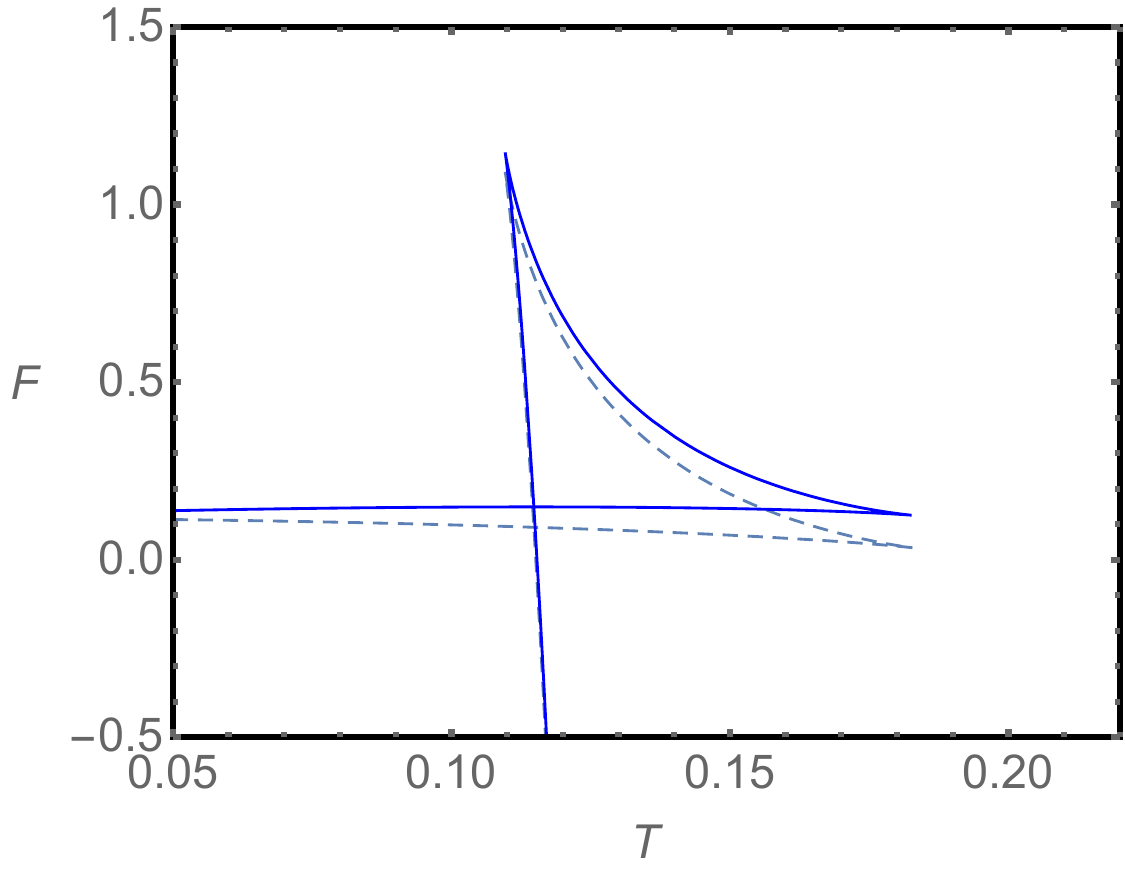}}\quad
    \subfigure[The quantum corrected free energy of EGB black hole in 6D \label{}]{\includegraphics[width=0.4\textwidth]{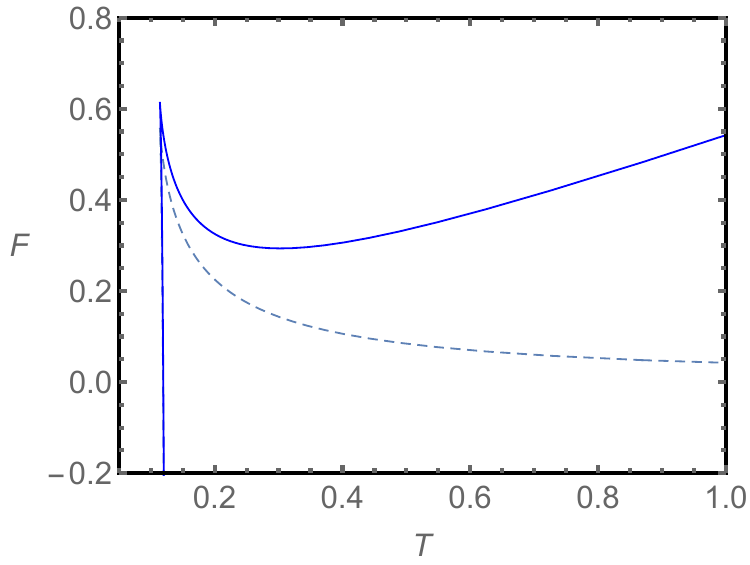}}\quad
  }
      \caption{The quantum corrected black hole free energy (solid blue line) is plotted along with the classical expression of free energy (dashed line).}
  \label{fig_3}
\end{figure*}
We interpret the ensemble-averaged free energy as the free energy of the classical solution, with additional contributions from other Euclidean geometries. According to the saddle-point approximation, and as explicitly shown in the previous section, the classical solution provides the dominant contribution, corresponding to small values of \( G_N \). Consequently, non-classical contributions become more significant at larger values of \( G_N \). To further investigate the effect of \( G_N \), we expand the free energy around the classical solution, i.e., around \( r_h = r_H \), where $r_H$ denotes the horizon radius of the saddle. This is justified when \( G_N \) is sufficiently small, as only the geometries near the saddle point contribute significantly to the free energy. To this extent, consider the parameter $\Delta = r_h-r_H$. Expanding the Euclidean action for EGB gravity, Eq. \eqref{IE_D} around $\Delta=0$ gives,
\beq
I_E^\Delta (r_h) \ \approx \  I_E(r_H)+ \frac{1}{2}\frac{d^2I_E}{dr_h^2}\Big|_{r_h=r_H} \Delta^2.
\eeq
The term linear in $\Delta$ vanishes as it corresponds to the extremum free energy. The expression for the generalized free energy is then given by,
\beq
F^\Delta (r_h) = \frac{I_E^\Delta (r_h) }{\beta}.
\eeq
Now, $e^{-I_E^\Delta }$ essentially becomes Gaussian distribution and one can use Gaussian integral to obtain the approximate value for the averaged-free energy, i.e.,
\beq
\ave{F} \ \approx \  \frac{\int_{-5\sigma}^{5\sigma} F^\Delta \ e^{-I_E^\Delta } d\Delta }{\int_{-5\sigma}^{5\sigma}  \ e^{-I_E^\Delta } d\Delta}.
\eeq
Here $\sigma$  is the standard deviation of the Gaussian distribution,
\beq
\sigma= \frac{1}{\sqrt{\frac{d^2I_E}{dr_h^2}}}.
\eeq
The integration yields,
\beq 
\ave{F} \ \approx \frac{I_E(r_H)}{\beta} + \frac{T_H}{2} -\frac{5 e^{-25/2}}{\sqrt{2 \pi } \text{erf}\left(\frac{5}{\sqrt{2}}\right)} T_H.
\eeq
The last term depends on the choice of integration limit and is neglected since it is very small compared to \( T_H / 2 \). The result above suggests that the ensemble-averaged free energy near the saddle geometry is given by the classical black hole free energy plus a correction term from non-classical geometries, i.e., \( T_H / 2 \). Note that this derivation applies to all black holes with asymptotic AdS behavior. This universal characteristic of free energy was reported in \cite{Cheng:2024efw}, and our results reinforce this claim by extending it to the case of higher-curvature gravity theories. A more realistic description of black hole thermodynamics for small values of \( G_N \) is obtained by defining a quantum-corrected free energy as follows:
\beq
F_{\text{corrected}} = F_{\text{classical}} + \frac{T_H}{2}.
\eeq
For the case of arbitrary dimensional EGB gravity theory, the expression for quantum corrected black hole free energy takes the form,
\begin{widetext}
\begin{eqnarray} \label{Q_free_D} \nonumber
F= \frac{1}{8\pi r_H (r_H^2+2 \tilde{\alpha})}\Bigg[(D-3) r_H^2 + \frac{D-1}{L^2} r_H^4+(D-5) \tilde{\alpha}-\frac{(D-3)}{r_H^{2(D-4)}} q^2\Bigg]+
\frac{(D-2)\Omega_{D-2}r_H^{D-3} }{16 \pi G_N}\Bigg[\left(1+\frac{r_H^2}{L^2}+\frac{\tilde{\alpha}}{r_H^2} +\frac{q^2}{r_H^{2(D-3)}}\right)\\ 
- \frac{1}{(r_H^2+2 \tilde{\alpha})}\left(1+\frac{2 \tilde{\alpha} (D-2)}{(D-4)}\frac{1}{r_H^2}\right)\left((D-3) r_H^2 + \frac{D-1}{L^2} r_H^4+(D-5) \tilde{\alpha}-(D-3) q^2 r_H^{-2(D-4)}\right)\Bigg].
\end{eqnarray}   
\end{widetext}

The quantum-corrected free energy for both five- and six-dimensional neutral cases is plotted alongside the corresponding classical expressions in Fig.~\ref{fig_3}. In the five-dimensional EGB case, the quantum-corrected free energy closely follows the behavior of the classical expression. However, in the six-dimensional case, an intriguing difference appears in the \( F\text{-}T \) diagram: the quantum-corrected free energy exhibits a local minimum, a feature absent in the classical free energy \cite{Cai:2013qga}.\\

Note that when \( G_N \to 0 \), only the classical geometry, which depends on \( 1/G_N \), contributes to the free energy. The sub-leading term is independent of \( G_N \). Also, as \( G_N \) increases, contributions from the higher powers of \( G_N \) become significant. Thus, one can express the ensemble-averaged free energy as a series in \( G_N \), i.e.,
\begin{equation} \label{series_F}
\langle F \rangle = \frac{F_0}{G_N} + F_1 \, G_N^0 + F_2 G_N^2 + \dots
\end{equation}
\begin{figure*}[hbt!]
    \centering
\mbox{
    \subfigure[ The leading order coefficient $F_0$ is shown in red.]{\includegraphics[width=0.4\textwidth]{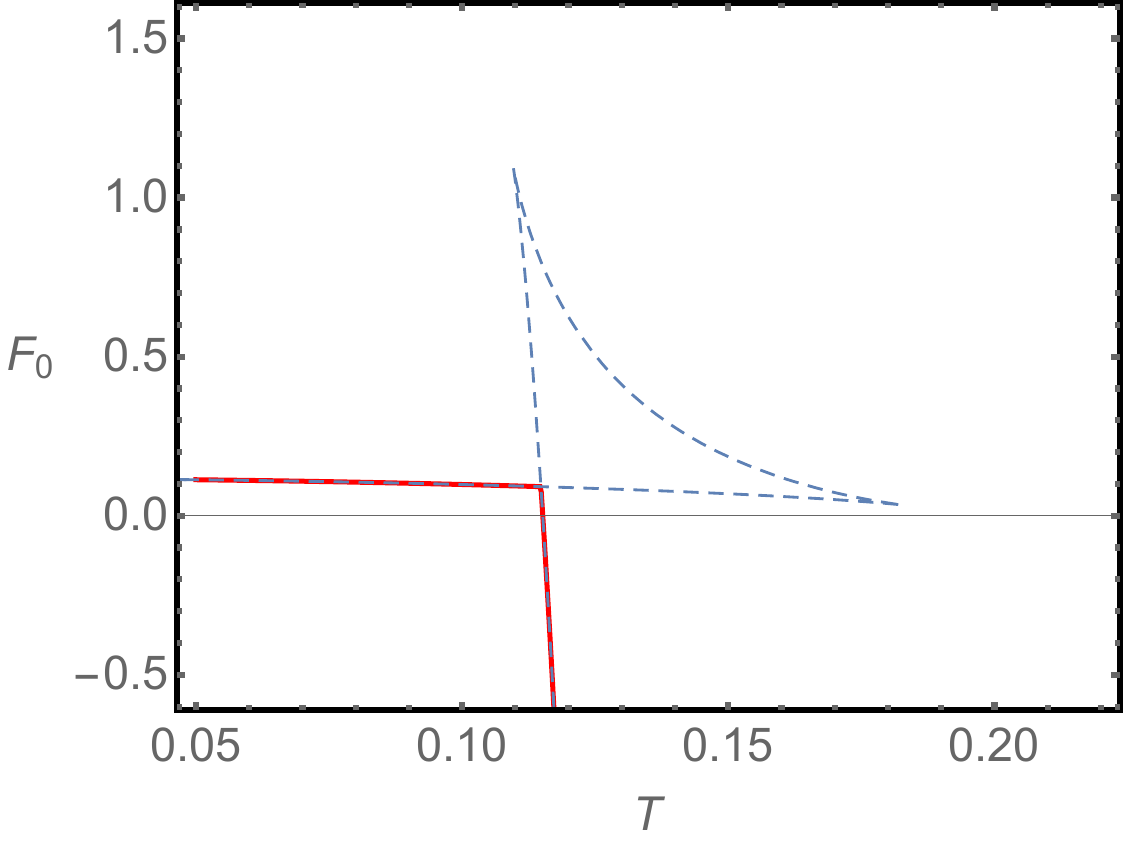}}\quad 
   
  }
  \mbox{
   \subfigure[The sub-leading coefficient $F_1$. ]{\includegraphics[width=0.4\textwidth]{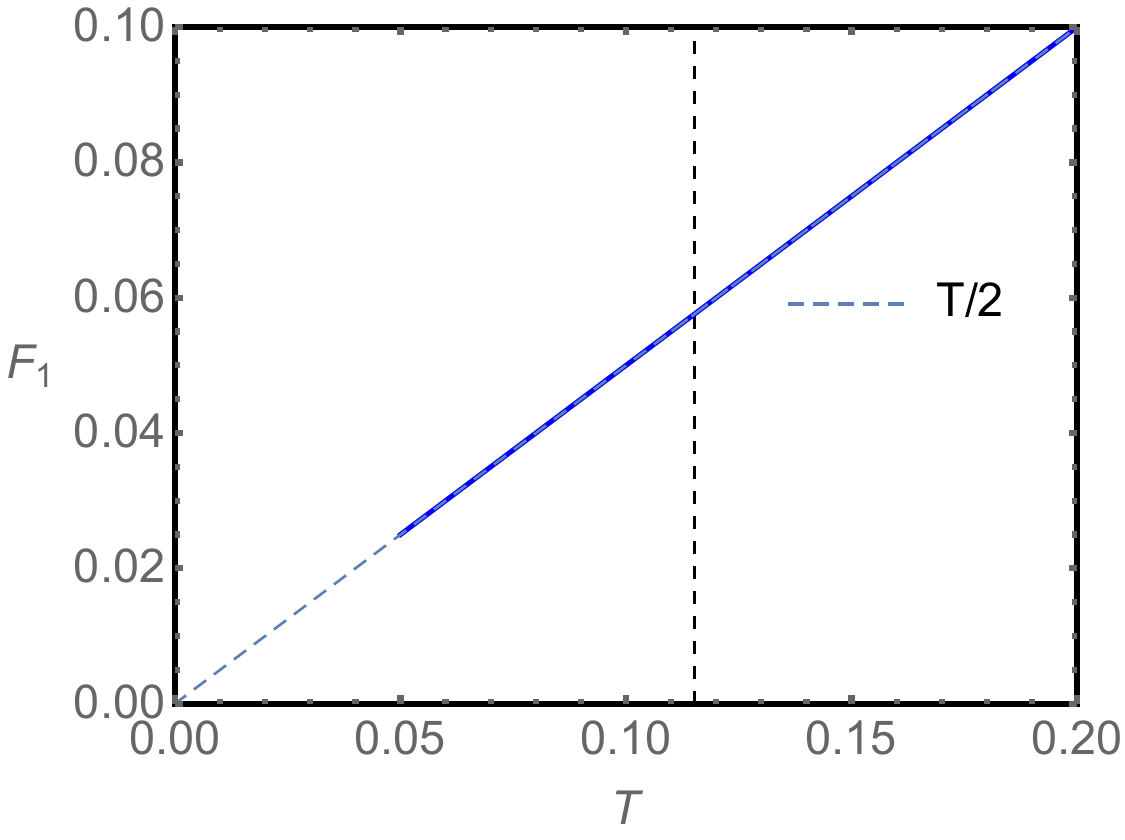}}\quad
    \subfigure[The sub-sub-leading coefficient $F_2$]{\includegraphics[width=0.4\textwidth]{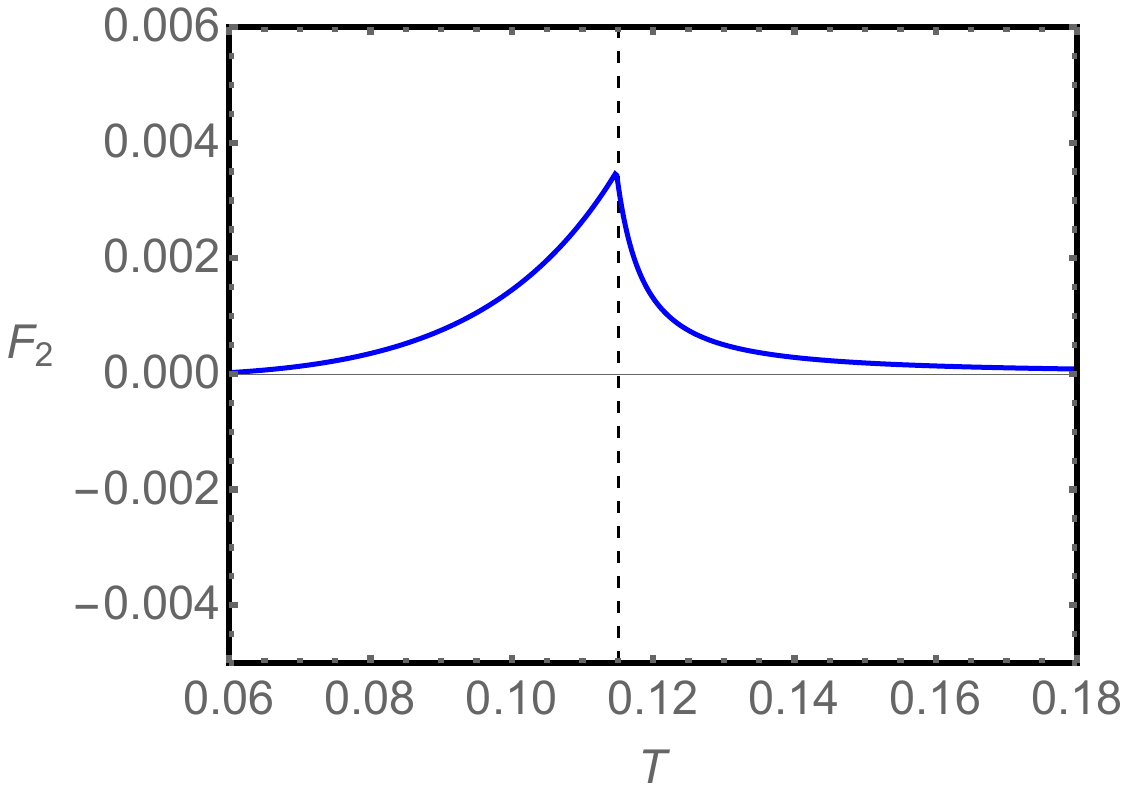}}\quad
    }
      \caption{(a) The leading order term matches the classical free energy for 5D EGB black hole. The dashed line represents $G_N F_{\text{classical}}$(b) The sub-leading term is shown in the solid blue curve and it matches with the analytical value(dashed line). The vertical line represents the classical transition temperature. (c) The sub-sub-leading order contribution to the averaged free energy is asymmetric to the classical transition temperature. }
  \label{fig_4}
\end{figure*}

In this spirit, one can verify the analytical results using numerical techniques. To this extent, we explicitly construct a data set for \( \langle F \rangle \) corresponding to various values of \( G_N \) and \( T \). This allows us to obtain the graphical representation of \( \langle F \rangle \) for a given value of \( G_N \), as shown in Figs.~\ref{fig_1} and~\ref{fig_2}. Further, we demonstrate the series expansion for \( \langle F \rangle \) by plotting the coefficients \( F_0 \), \( F_1 \), and \( F_2 \) in Eq.~\eqref{series_F} using the method of curve fitting.\\

The averaged free energy obtained from numerical calculations for the five-dimensional EGB black hole is summarized in Fig. \ref{fig_4}. For $G_N=1/1000$, the coefficients $F_0, F_1$, and $F_2$ are plotted against ensemble temperature. The numerical results match the analytical expressions for leading and sub-leading order contributions. Our results are qualitatively similar to the case of RN-AdS and Kerr-AdS black holes \cite{Cheng:2024efw,Cheng:2024hxh}.\\

\section{discussion}\label{dis}

This work introduces an ensemble-averaged free energy for black holes in EGB gravity theory. In the standard free energy landscape description of black hole thermodynamics, the generalized free energy is derived from the gravitational partition function using the saddle-point approximation. However, a more physically relevant expression for the free energy, called the ensemble-averaged free energy, can be obtained by including non-saddle geometries in the integration, as outlined in \cite{Cheng:2024efw}. Using this approach, we analyze the thermodynamic properties of EGB black holes through the ensemble-averaged free energy. The ensemble-averaged free energy does not display a sharp transition point, suggesting that the black hole phase transition can be viewed as the small \( G_N \) approximation within the ensemble-averaged theory. \\

The ensemble-averaged free energy can be expanded in powers of \( G_N \), with the leading order term corresponding to the classical black hole free energy. For both five- and six-dimensional EGB black holes, the sub-leading term is \( T_H/2 \). Our results support the conjecture that this sub-leading contribution remains \( T_H/2 \) regardless of the specific gravity theory \cite{Cheng:2024efw,Cheng:2024hxh}. Additionally, we derive an expression for the quantum-corrected free energy for EGB black holes. In the case of a six-dimensional EGB black hole, the quantum-corrected free energy exhibits a distinct local minimum, a feature not present in the classical free energy. This feature is also observed in the Schwarzschild-AdS black hole. Additionally, the ensemble-averaged free energy structure is identical for both of these black holes. Thus, we conclude that the striking similarities in thermodynamic behavior between the Schwarzschild-AdS and six-dimensional EGB-AdS black holes and between the RN-AdS and five-dimensional EGB-AdS black holes extend beyond the classical limit.

\section{Acknowledgements}
This work was supported by the National Natural Science Foundation of China, (Grant No.12347177, and No. 12405073). The authors would like to thank Soumen Roy for his assistance with the numerical calculations.

\appendix

\bibliography{BibTex}

\providecommand{\latin}[1]{#1}
\makeatletter
\providecommand{\doi}
  {\begingroup\let\do\@makeother\dospecials
  \catcode`\{=1 \catcode`\}=2 \doi@aux}
\providecommand{\doi@aux}[1]{\endgroup\texttt{#1}}
\makeatother
\providecommand*\mcitethebibliography{\thebibliography}
\csname @ifundefined\endcsname{endmcitethebibliography}
  {\let\endmcitethebibliography\endthebibliography}{}
\begin{mcitethebibliography}{16}
\providecommand*\natexlab[1]{#1}
\providecommand*\mciteSetBstSublistMode[1]{}
\providecommand*\mciteSetBstMaxWidthForm[2]{}
\providecommand*\mciteBstWouldAddEndPuncttrue
  {\def\EndOfBibitem{\unskip.}}
\providecommand*\mciteBstWouldAddEndPunctfalse
  {\let\EndOfBibitem\relax}
\providecommand*\mciteSetBstMidEndSepPunct[3]{}
\providecommand*\mciteSetBstSublistLabelBeginEnd[3]{}
\providecommand*\EndOfBibitem{}
\mciteSetBstSublistMode{f}
\mciteSetBstMaxWidthForm{subitem}{(\alph{mcitesubitemcount})}
\mciteSetBstSublistLabelBeginEnd
  {\mcitemaxwidthsubitemform\space}
  {\relax}
  {\relax}

\bibitem[Hawking(1983)]{Hawking:sw}
Hawking,~D.,~S.W.and~Page {Thermodynamics of black holes in anti-de Sitter
  space}. \emph{Commun.Math. Phys.} \textbf{1983}, \emph{87}, 577–588\relax
\mciteBstWouldAddEndPuncttrue
\mciteSetBstMidEndSepPunct{\mcitedefaultmidpunct}
{\mcitedefaultendpunct}{\mcitedefaultseppunct}\relax
\EndOfBibitem
\bibitem[Witten(1998)]{Witten:1998zw}
Witten,~E. {Anti-de Sitter space, thermal phase transition, and confinement in
  gauge theories}. \emph{Adv. Theor. Math. Phys.} \textbf{1998}, \emph{2},
  505--532\relax
\mciteBstWouldAddEndPuncttrue
\mciteSetBstMidEndSepPunct{\mcitedefaultmidpunct}
{\mcitedefaultendpunct}{\mcitedefaultseppunct}\relax
\EndOfBibitem
\bibitem[Kubiznak \latin{et~al.}(2017)Kubiznak, Mann, and
  Teo]{Kubiznak:2016qmn}
Kubiznak,~D.; Mann,~R.~B.; Teo,~M. {Black hole chemistry: thermodynamics with
  Lambda}. \emph{Class. Quant. Grav.} \textbf{2017}, \emph{34}, 063001\relax
\mciteBstWouldAddEndPuncttrue
\mciteSetBstMidEndSepPunct{\mcitedefaultmidpunct}
{\mcitedefaultendpunct}{\mcitedefaultseppunct}\relax
\EndOfBibitem
\bibitem[Maldacena(1998)]{Maldacena:1997re}
Maldacena,~J.~M. {The Large N limit of superconformal field theories and
  supergravity}. \emph{Adv. Theor. Math. Phys.} \textbf{1998}, \emph{2},
  231--252\relax
\mciteBstWouldAddEndPuncttrue
\mciteSetBstMidEndSepPunct{\mcitedefaultmidpunct}
{\mcitedefaultendpunct}{\mcitedefaultseppunct}\relax
\EndOfBibitem
\bibitem[Witten(1998)]{Witten:1998qj}
Witten,~E. {Anti-de Sitter space and holography}. \emph{Adv. Theor. Math.
  Phys.} \textbf{1998}, \emph{2}, 253--291\relax
\mciteBstWouldAddEndPuncttrue
\mciteSetBstMidEndSepPunct{\mcitedefaultmidpunct}
{\mcitedefaultendpunct}{\mcitedefaultseppunct}\relax
\EndOfBibitem
\bibitem[Kovtun \latin{et~al.}(2005)Kovtun, Son, and Starinets]{Kovtun:2004de}
Kovtun,~P.; Son,~D.~T.; Starinets,~A.~O. {Viscosity in strongly interacting
  quantum field theories from black hole physics}. \emph{Phys. Rev. Lett.}
  \textbf{2005}, \emph{94}, 111601\relax
\mciteBstWouldAddEndPuncttrue
\mciteSetBstMidEndSepPunct{\mcitedefaultmidpunct}
{\mcitedefaultendpunct}{\mcitedefaultseppunct}\relax
\EndOfBibitem
\bibitem[Hartnoll \latin{et~al.}(2008)Hartnoll, Herzog, and
  Horowitz]{Hartnoll:2008vx}
Hartnoll,~S.~A.; Herzog,~C.~P.; Horowitz,~G.~T. {Building a Holographic
  Superconductor}. \emph{Phys. Rev. Lett.} \textbf{2008}, \emph{101},
  031601\relax
\mciteBstWouldAddEndPuncttrue
\mciteSetBstMidEndSepPunct{\mcitedefaultmidpunct}
{\mcitedefaultendpunct}{\mcitedefaultseppunct}\relax
\EndOfBibitem
\bibitem[Hartnoll \latin{et~al.}(2007)Hartnoll, Kovtun, Muller, and
  Sachdev]{Hartnoll:2007ih}
Hartnoll,~S.~A.; Kovtun,~P.~K.; Muller,~M.; Sachdev,~S. {Theory of the Nernst
  effect near quantum phase transitions in condensed matter, and in dyonic
  black holes}. \emph{Phys. Rev. B} \textbf{2007}, \emph{76}, 144502\relax
\mciteBstWouldAddEndPuncttrue
\mciteSetBstMidEndSepPunct{\mcitedefaultmidpunct}
{\mcitedefaultendpunct}{\mcitedefaultseppunct}\relax
\EndOfBibitem
\bibitem[Gibbons and Hawking(1977)Gibbons, and Hawking]{Gibbons:1976ue}
Gibbons,~G.~W.; Hawking,~S.~W. {Action Integrals and Partition Functions in
  Quantum Gravity}. \emph{Phys. Rev. D} \textbf{1977}, \emph{15},
  2752--2756\relax
\mciteBstWouldAddEndPuncttrue
\mciteSetBstMidEndSepPunct{\mcitedefaultmidpunct}
{\mcitedefaultendpunct}{\mcitedefaultseppunct}\relax
\EndOfBibitem
\bibitem[Gibbons \latin{et~al.}(1978)Gibbons, Hawking, and
  Perry]{Gibbons:1978ac}
Gibbons,~G.~W.; Hawking,~S.~W.; Perry,~M.~J. {Path Integrals and the
  Indefiniteness of the Gravitational Action}. \emph{Nucl. Phys. B}
  \textbf{1978}, \emph{138}, 141--150\relax
\mciteBstWouldAddEndPuncttrue
\mciteSetBstMidEndSepPunct{\mcitedefaultmidpunct}
{\mcitedefaultendpunct}{\mcitedefaultseppunct}\relax
\EndOfBibitem
\bibitem[Hawking(1978)]{Hawking:1978jz}
Hawking,~S.~W. {Quantum Gravity and Path Integrals}. \emph{Phys. Rev. D}
  \textbf{1978}, \emph{18}, 1747--1753\relax
\mciteBstWouldAddEndPuncttrue
\mciteSetBstMidEndSepPunct{\mcitedefaultmidpunct}
{\mcitedefaultendpunct}{\mcitedefaultseppunct}\relax
\EndOfBibitem
\bibitem[Cheng \latin{et~al.}(2024)Cheng, Pan, Xu, and Yang]{Cheng:2024efw}
Cheng,~P.; Pan,~J.; Xu,~H.; Yang,~S.-J. {Thermodynamics of the Kerr-AdS black
  hole from an ensemble-averaged theory}. \textbf{2024}, \relax
\mciteBstWouldAddEndPunctfalse
\mciteSetBstMidEndSepPunct{\mcitedefaultmidpunct}
{}{\mcitedefaultseppunct}\relax
\EndOfBibitem
\bibitem[Cheng \latin{et~al.}(2024)Cheng, Liu, and Wei]{Cheng:2024hxh}
Cheng,~P.; Liu,~Y.-X.; Wei,~S.-W. {Black hole thermodynamics from an
  ensemble-averaged theory}. \textbf{2024}, \relax
\mciteBstWouldAddEndPunctfalse
\mciteSetBstMidEndSepPunct{\mcitedefaultmidpunct}
{}{\mcitedefaultseppunct}\relax
\EndOfBibitem
\bibitem[Lovelock(1971)]{Lovelock:1971yv}
Lovelock,~D. {The Einstein tensor and its generalizations}. \emph{J. Math.
  Phys.} \textbf{1971}, \emph{12}, 498--501\relax
\mciteBstWouldAddEndPuncttrue
\mciteSetBstMidEndSepPunct{\mcitedefaultmidpunct}
{\mcitedefaultendpunct}{\mcitedefaultseppunct}\relax
\EndOfBibitem
\bibitem[Cai \latin{et~al.}(2013)Cai, Cao, Li, and Yang]{Cai:2013qga}
Cai,~R.-G.; Cao,~L.-M.; Li,~L.; Yang,~R.-Q. {P-V criticality in the extended
  phase space of Gauss-Bonnet black holes in AdS space}. \emph{JHEP}
  \textbf{2013}, \emph{09}, 005\relax
\mciteBstWouldAddEndPuncttrue
\mciteSetBstMidEndSepPunct{\mcitedefaultmidpunct}
{\mcitedefaultendpunct}{\mcitedefaultseppunct}\relax
\EndOfBibitem
\bibitem[Li and Wang(2023)Li, and Wang]{Li:2023men}
Li,~R.; Wang,~J. {Generalized free energy landscapes of charged
  Gauss-Bonnet-AdS black holes in diverse dimensions}. \emph{Phys. Rev. D}
  \textbf{2023}, \emph{108}, 044057\relax
\mciteBstWouldAddEndPuncttrue
\mciteSetBstMidEndSepPunct{\mcitedefaultmidpunct}
{\mcitedefaultendpunct}{\mcitedefaultseppunct}\relax
\EndOfBibitem
\end{mcitethebibliography}


\end{document}